\documentstyle[psfig,prb,twocolumn,aps]{revtex}
\begin{document}
\draft

\twocolumn[\hsize\textwidth\columnwidth\hsize\csname@twocolumnfalse\endcsname

\title{Thermodynamics of the spin-flop transition in a quantum $XYZ$ chain}

\author{X. Wang and X. Zotos}
\address{Institut Romand de Recherche Num\'erique en
Physique des Mat\'eriaux (IRRMA), PPH-Ecublens, CH-1015 Lausanne, Switzerland}

\author{J. Karadamoglou and N. Papanicolaou }
\address{Department of Physics, University of Crete, and 
Research Center of Crete, Heraklion, Greece}
\maketitle

\begin{abstract}
A special limit of an antiferromagnetic $XYZ$ chain was
recently shown to exhibit interesting bulk as well as surface
spin-flop transitions at $T=0$. Here we provide a complete calculation
of the thermodynamics of the bulk transition using a
 transfer-matrix-renormalization-group (TMRG) method that addresses
directly the thermodynamic limit of quantum spin chains. We also 
shed some light on certain spinwave anomalies at low temperature
predicted earlier by Johnson and Bonner.
\end{abstract}

]
There has been a revival of interest in bulk and surface
spin-flop transitions following some recent experimental work on
Fe/Cr multilayers.\cite{1} These are effectively described by classical
spin chains characterized by antiferromagnetic exchange interaction
in addition to single ion anisotropy. It is then natural to raise
similar questions in the context of quantum spin chains which are 
more appropriate for the study of quasi-one-dimensional crystalline
magnetic systems.

Indeed, in a recent communication,\cite{2} this issue was studied within
a special limit of the spin$-{1\over 2}$ $XYZ$ chain defined by 
the Hamiltonian
\begin{eqnarray}
\label{eq:1}
	W 
	=
	&& 
	- \sum_{\ell=1}^{\Lambda}
		\left[ 
			T_{\ell}^x T_{\ell+1}^x  +
			T_{\ell}^y T_{\ell+1}^y  + \Delta 
			\left( 
				T_{\ell}^z T_{\ell+1}^z - {1\over 4}
			\right) \right.
	\nonumber
	\\
	&&
	\left. + H (-1)^{\ell} T_{\ell}^z \right] ,
\end{eqnarray}
where $\Delta >1$ and the operators $\bf T_{\ell}$ satisfy the
standard spin commutation relations; hence (\ref{eq:1}) may be thought of 
as the Hamiltonian of a ferromagnetic $XXZ$ chain in a staggered 
magnetic field. A more physical interpretation is obtained by the
canonical transformation $S^x_{\ell}=T^x_{\ell}$,
$S^y_{\ell}=(-1)^{\ell}T^y_{\ell}$, $S^z_{\ell}=(-1)^{\ell}T^z_{\ell}$
which reduces (\ref{eq:1}) to a special limit of an antiferromagnetic $XYZ$
Hamiltonian in a {\sl uniform} field $H$. In view of the above dual 
interpretation we consider in parallel the two special operators
\begin{equation}
\label{eq:2}
\tau =\sum^{\Lambda}_{\ell =1}T^z_{\ell},\qquad
M=\sum^{\Lambda}_{\ell =1}(-1)^{\ell}T^z_{\ell}.
\end{equation}
The operator $\tau$ is not endowed with a simple physical meaning 
but commutes with Hamiltonian (\ref{eq:1}) and thus provides a very useful 
classification of states. In contrast, the operator $M$ does not 
commute with the Hamiltonian but represents the physical 
magnetization, a quantity of special interest in the following.

Although the main objective of this paper is to study the 
thermodynamic limit $\Lambda\to\infty$, some issues become clear
by approaching that limit through a finite periodic chain with 
an even number of sites $\Lambda =2N$. The eigenvalues of $\tau$
are then given by $\tau =0, \pm 1, \ldots , \pm N$ and split the 
Hilbert space into $2N+1$ sectors. The two extremal sectors
$\tau =\pm N$ contain only one state each, which is an exact 
eigenstate of the Hamiltonian with energy $E=0$ for any strength
of the applied field $H$. In the original spin language these are 
the two completely polarized N\'{e}el states. To study their 
stability at finite $H$ we also consider one-magnon excitations,
with $\tau =-N+1$ or $N-1$, whose energy eigenvalues are the 
same for both sectors and are given explicitly by
\begin{equation}
\label{eq:3}
E_k=\Delta\pm\sqrt{\cos^2(k/2)+H^2},\qquad
G_{\pm}=\Delta \pm\sqrt{1+H^2},
\end{equation}
where $k$ is a sublattice crystal momentum. In Eq. (\ref{eq:3}) we also 
display the energies of the $k=0$ modes, denoted by $G_{\pm}$,
which will be referred to as the magnon gaps and are depicted in 
Fig. 1 as functions of the applied field. 
\begin{figure}

\centerline{\hbox{\psfig{figure=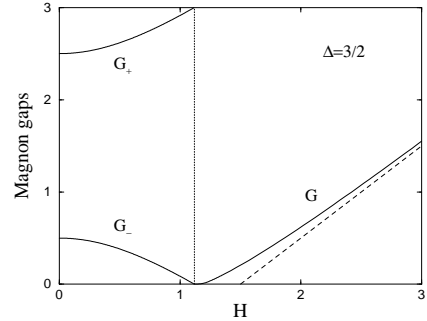,width=5.5cm}}}

\caption{Field dependence of the magnon gaps for the specific
anisotropy $\Delta =3/2$ for which the critical field is 
$H_b=1.118$. The skew dashed line represents the Ising asymptote
$H-\Delta$.}
\end{figure}
It is clear that the 
lowest gap closes $(G_-=0)$ at the critical field
\begin{equation}
\label{eq:4}
H_b=\sqrt{\Delta^2-1}
\end{equation}
beyond which the N\'{e}el states are no longer the lowest-energy
states and the system undergoes a bulk spin-flop (BSF) transition.

The nature of this $T=0$ phase transition is actually more 
interesting than indicated by the preceding argument. In fact, the 
lowest-energy states of {\sl all} sectors become degenerate at the 
critical field, with energy $E=0$, and the corresponding eigenstates
can be constructed analytically.\cite{2} For $H> H_b$ the $\tau =0$
sector prevails in the sense that it contains the unique absolute 
ground state. Accordingly the first 
excited states are the lowest-energy states of the $\tau =\pm 1$
sectors and are degenerate. The corresponding magnon gap, denoted
by $G$ in Fig. 1, was computed via a Lanczos algorithm, on finite   
periodic chains with $\Lambda\le 22$, complemented by 
straightforward Richardson extrapolation.\cite{3} The detailed
numerical results indicate that the gap $G$ might vanish through
an essential singularity at $H_b$ in the thermodynamic limit.
Extrapolation becomes completely unnecessary for fields in the 
region $H\gtrsim\Delta$ where the gap $G$ approaches the Ising 
asymptote $H-\Delta$. One would expect that the lowest gaps 
$G_-$ and $G$ dominate the low-temperature thermodynamics,
in the respective field ranges, an issue that turned out to be 
more intricate than normally anticipated.

In order to prepare the discussion of thermodynamics it is also 
useful to calculate the magnetization $M$ at $T=0$. The 
magnetization vanishes for $H< H_b$ but exhibits a finite jump 
at the critical field which can be calculated analytically. For
$H=H_b$ the expected value of $M$ in the ground state  
$|\psi_{\tau}\rangle$ of each sector $\tau$ is 
\begin{eqnarray}
\label{eq:5}
	&&
	M_{\tau}
	=
	\,{\langle\psi_{\tau}|M|\psi_{\tau}\rangle
	\over\langle\psi_{\tau}|\psi_{\tau}\rangle}=\sqrt{\Delta^2-1}
	{NI_{_{N-1}}\over I_{_N}},
	\nonumber
	\\
	&&
	I_{_N}=\,{1\over\pi}\int^{\pi}_0\cos (\tau\theta)
	(\Delta +\cos\theta)^{^N}d\theta ,
\end{eqnarray}
which generalizes the $\tau =0$ result quoted in Ref. 2. For any 
fixed $\tau$ a simple application of the Laplace method \cite{3} yields
the asymptotic expansion
\begin{equation}
\label{eq:6}
M_{\tau}=\left (N+{1\over 2}\right)
\sqrt{{\Delta -1\over\Delta +1}}+{1-4\tau^2\over 8N}
\sqrt{\Delta^2-1}+O(1/N^2),
\end{equation}
which can be used to extract the thermodynamic limit. Since the 
$\tau =0$ sector contains the absolute ground state just above
$H_b$, the magnetization jump at the critical field is given by
\begin{equation}
\label{eq:7}
\mu_0=\lim_{N\to\infty}\left({M_0\over 2N}\right)=
{1\over 2}\sqrt{{\Delta -1\over\Delta +1}}\, .
\end{equation}
A more subtle quantity is the $T=0$ {\sl average} magnetization 
at the critical point calculated from
\begin{equation}
\label{eq:8}
\mu_b=\lim_{N\to\infty}{1\over (2N)^2}\sum^N_{\tau =-N}M_{\tau}.
\end{equation}
The asymptotic expansion (\ref{eq:6}) cannot be employed in Eq. (\ref{eq:8}) because
the latter contains terms with values of $\tau$ that are comparable
to $N$. Nevertheless the explicit result (\ref{eq:5}) may be inserted in 
Eq. (\ref{eq:8}) to numerically estimate $\mu_b$ at large $N$.

For $H> H_b$ the $T=0$ magnetization is not known analytically and 
we have again resorted to the Lanczos algorithm. At our maximum 
size $\Lambda =22$ the Lanczos result for the magnetization jump 
at the critical field differs from the analytical prediction (\ref{eq:7})
by about 5\%, a difference that is rectified by Richardson 
extrapolation to an accuracy about one part in a thousand. Hence we 
have applied the same extrapolation for $H> H_b$ and the result is 
depicted by a solid line in Fig. 2. Again, extrapolation becomes 
unnecessary for $H\gtrsim\Delta$.

\begin{figure}

\centerline{\hbox{\psfig{figure=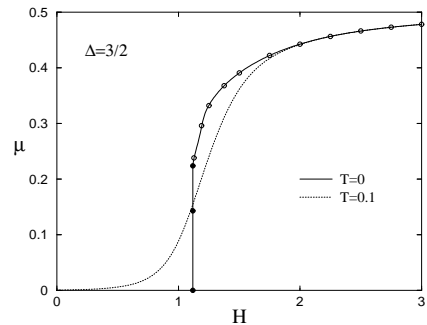,width=5.5cm}}}

\caption{Field dependence of the average magnetization per site
$\mu =M/\Lambda$ for $T=0$ (solid line) and $T=0.1$ (dotted line).
The solid circles represent the values of the $T=0$ magnetization 
just below $H_b$ $(\mu =0)$, right at $H_b$ $(\mu_b=0.143)$, and 
just above $H_b$ $(\mu_0=0.224)$. The meaning of the open circles 
is discussed in the text.}
\end{figure}

We thus arrive at the main point of this paper, namely the 
calculation of thermodynamics via a TMRG algorithm \cite{4} which has 
already been applied to the study of quantum spin ladders \cite{5}.
One of the distinct features of the method is that it directly
addresses the thermodynamic limit $\Lambda\to\infty$. We shall not
present here numerical details but merely discuss some important
results.

For instance, the temperature dependence of the magnetization is 
shown in Fig. 3 for a number of field values. 
The magnetization 
vanishes for all temperatures at vanishing field. For finite fields
in the subcritical region, $H< H_b$, the magnetization again vanishes 
at $T=0$, as expected, but develops a maximum at some finite
temperature. Right at the critical field, $H=H_b$, the $T=0$
limit of the calculated curve is consistent with the value 
$\mu_b=0.143$ of Eq. (\ref{eq:8}), applied for $\Delta =3/2$, whereas just 
above $H_b$ the $T=0$ limit is consistent with the value
$\mu_0=0.224$ of Eq. (\ref{eq:7}). For supercritical fields, $H> H_b$,
the low-temperature limiting values of $\mu$ extracted from Fig. 3
are depicted by open circles in Fig. 2 and are thus seen to be in 
excellent agreement with our independent Lanczos calculation of 
the magnetization at $T=0$. 
Similarly the results extracted from 
Fig. 3 at the specific temperature $T=0.1$ were used to calculate
the field dependence of the magnetization at this temperature, a 
result that is shown by a dotted line in Fig. 2 and illustrates the 
manner in which the $T=0$ magnetization jump at the critical point 
is smoothed out at finite temperature.
\begin{figure}

\centerline{\hbox{\psfig{figure=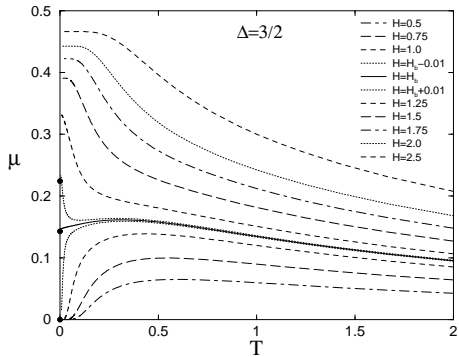,width=6.0cm}}}

\caption{Temperature dependence of the magnetization for 
various field values throughout the BSF transition. The solid circles 
on the $\mu$ axis correspond to the same values as those of Fig. 2.}
\end{figure}

We next turn our attention to the specific heat. In Fig. 4 we 
compare the TMRG result at vanishing field with a finite-size 
calculation for chains with $\Lambda =10, 12$, and 14 for which 
a complete numerical diagonalization of the Hamiltonian is possible.
This comparison is surprising in that the trend of the finite-size 
results does not seem to be consistent with the calculated 
thermodynamic limit. We have thus naturally questioned the validity
of our TMRG calculation. However this special case was also 
considered in Fig. 4b of a paper by Kl\"{u}mper \cite{6} whose numerical
method is again based on a transfer matrix but relies heavily on the 
complete integrability of model (\ref{eq:1}) at vanishing staggered field.
Direct correspondence with the above author established that our 
result agrees with his throughout the temperature range considered.
\begin{figure}

\centerline{\hbox{\psfig{figure=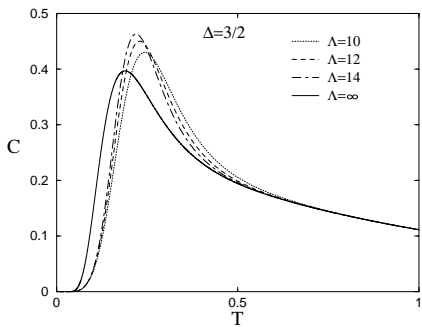,width=5.5cm}}}

\caption{Temperature dependence of the specific heat 
per site $C$ at vanishing field $(H=0)$.}
\end{figure}
It should be added here that the TMRG method does not rely on 
complete integrability and is thus more flexible; e.g., it can be 
applied for the calculation of the thermodynamics at any finite staggered
field for which model (\ref{eq:1}) is not known to be completely integrable.

The ``anomalous scaling'' observed in Fig. 4 for $H=0$ persists
for nonvanishing fields throughout the BSF transition but
gradually disappears in the ``no-scaling region'' $H\gtrsim\Delta$
where the correct thermodynamic limit is practically reached by 
very short chains, as short as $\Lambda =4$. In any case, the TMRG
calculation of the temperature dependence of the specific heat is 
illustrated in Fig. 5 for various field values. The main feature of 
this figure is that the specific heat develops a double peak for 
fields in the vicinity of the critical point $H_b$. Furthermore the 
low-temperature behavior appears to be generally consistent with the 
field dependence of the magnon gaps shown in Fig. 1.
\vspace*{-0.15cm}
\begin{figure}

\centerline{\hbox{\psfig{figure=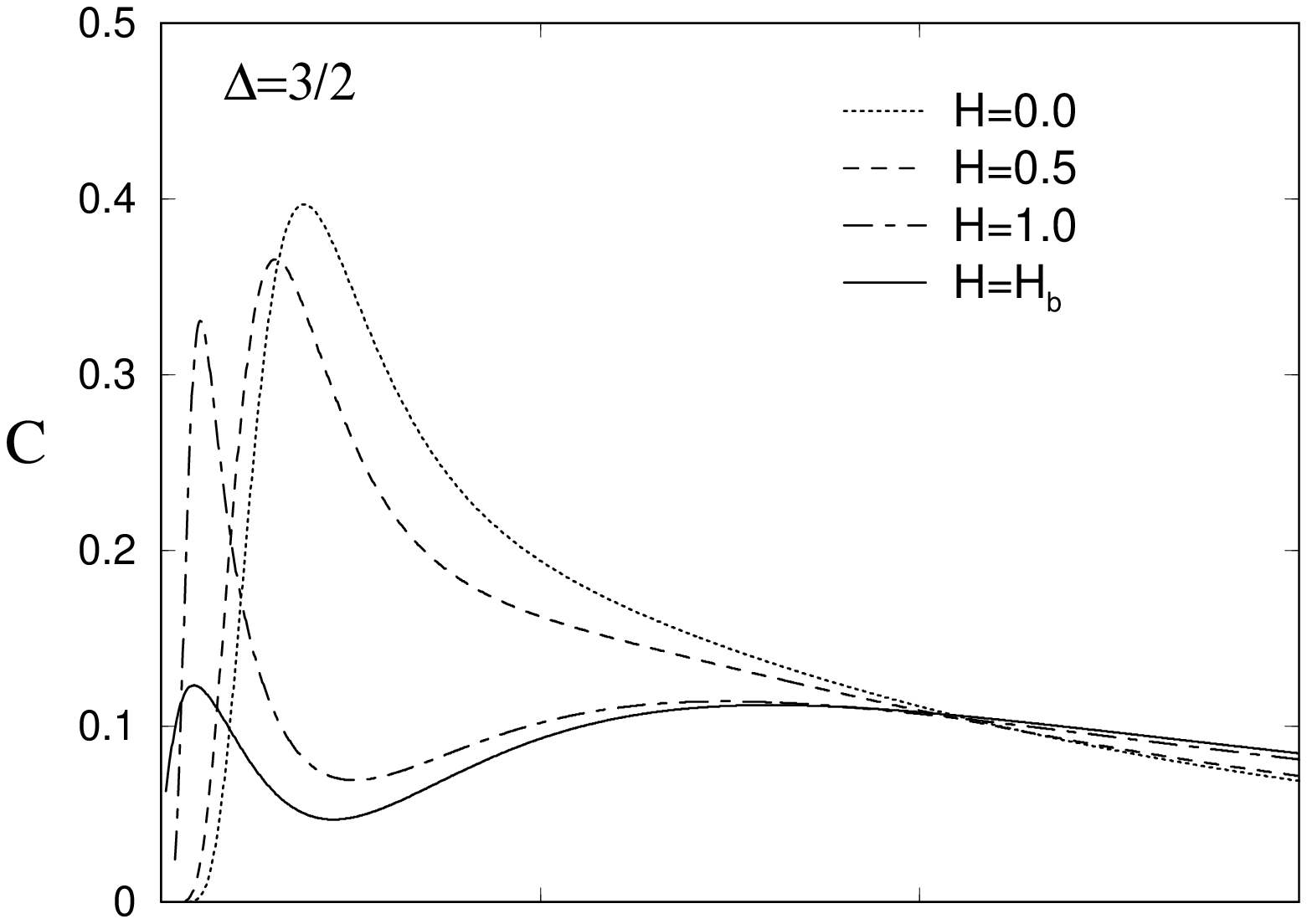,width=5.5cm}}}
\centerline{\hbox{\psfig{figure=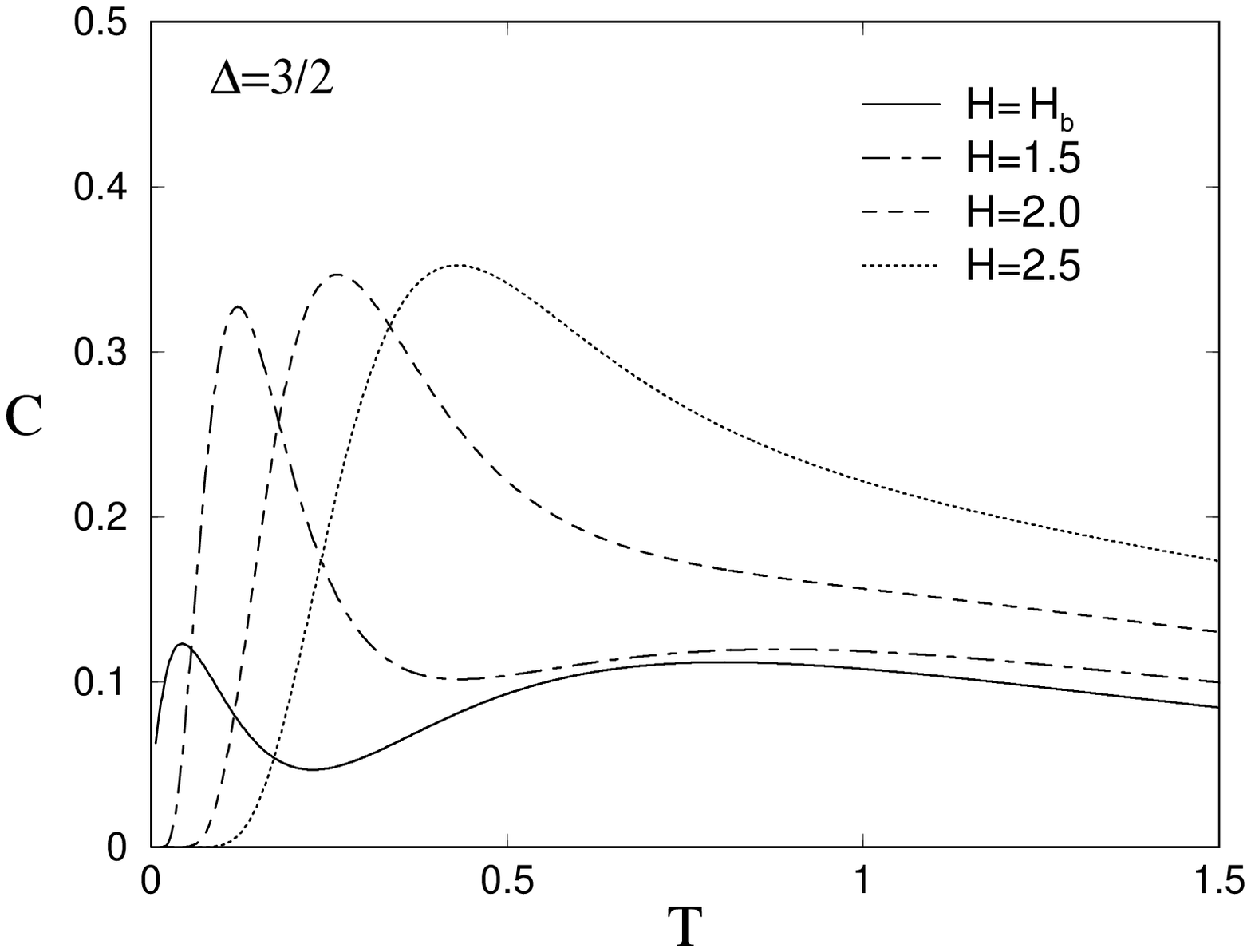,width=5.5cm}}}

\caption{Specific heat for various field values throughout 
the BSF transition.}
\end{figure}

One would expect that the low-temperature specific heat is correctly
predicted by a dilute-magnon or spinwave approximation, a long
cherished assumption in condensed matter physics. To check this 
assumption we first consider the case of vanishing field for which
our model is formally identical to the ferromagnetic $XXZ$ chain
extensively studied through the Bethe Ansatz \cite{6,7}. At sufficiently
low temperature the spinwave approximation of the specific heat 
should read 
\begin{equation}
\label{eq:9}
C\approx {G_-^2\exp (-G_-/T)\over (2\pi T^3)^{1/2}},\qquad
G_-=\Delta -1,
\end{equation}
where $G_-$ is the lowest magnon gap at vanishing field. Equation 
(\ref{eq:9}) suggests considering the quantity $-T\ln (T^{3/2}C)$ which
should interpolate linearly to the magnon gap $G_-$ at $T=0$. Yet a 
comparison of the spinwave prediction (\ref{eq:9}) with the TMRG 
calculation shown in the $\Delta =3/2$ entry of Fig. 6 reveals
a sharp disagreement even at the lowest temperature accessible
by our method. On the other hand, one can show that the spinwave
approximation agrees well with the finite-size results for 
$\Lambda =10, 12, 14$ given in Fig. 4 restricted to the temperature
range of Fig. 6. Clearly then the anomalous scaling noted earlier 
is intimately related to the apparent failure of the dilute-magnon
approximation.

\begin{figure}

\centerline{\hbox{\psfig{figure=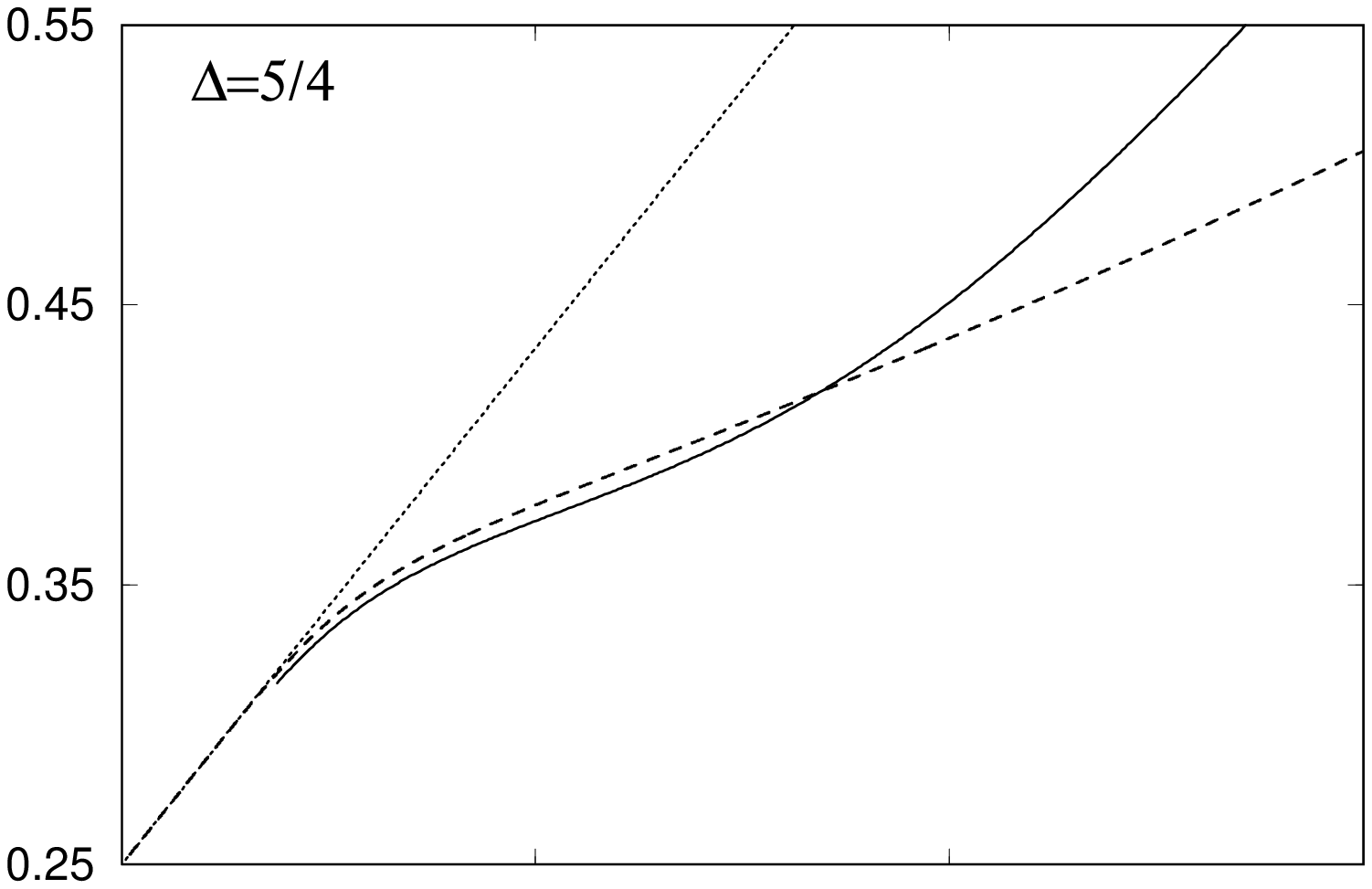,width=5.4cm}}}

\centerline{\hbox{\psfig{figure=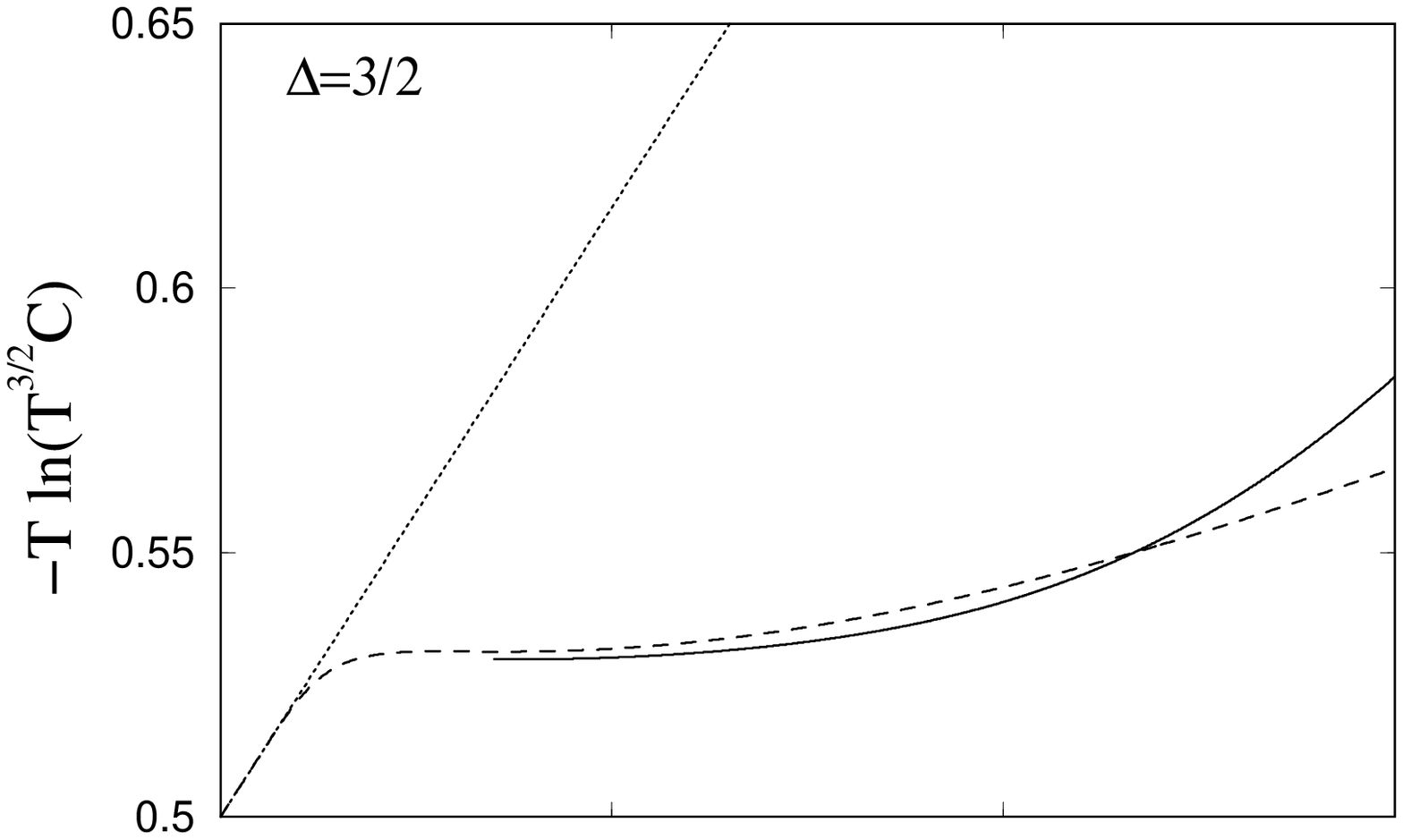,width=5.4cm}}}

\centerline{\hbox{\psfig{figure=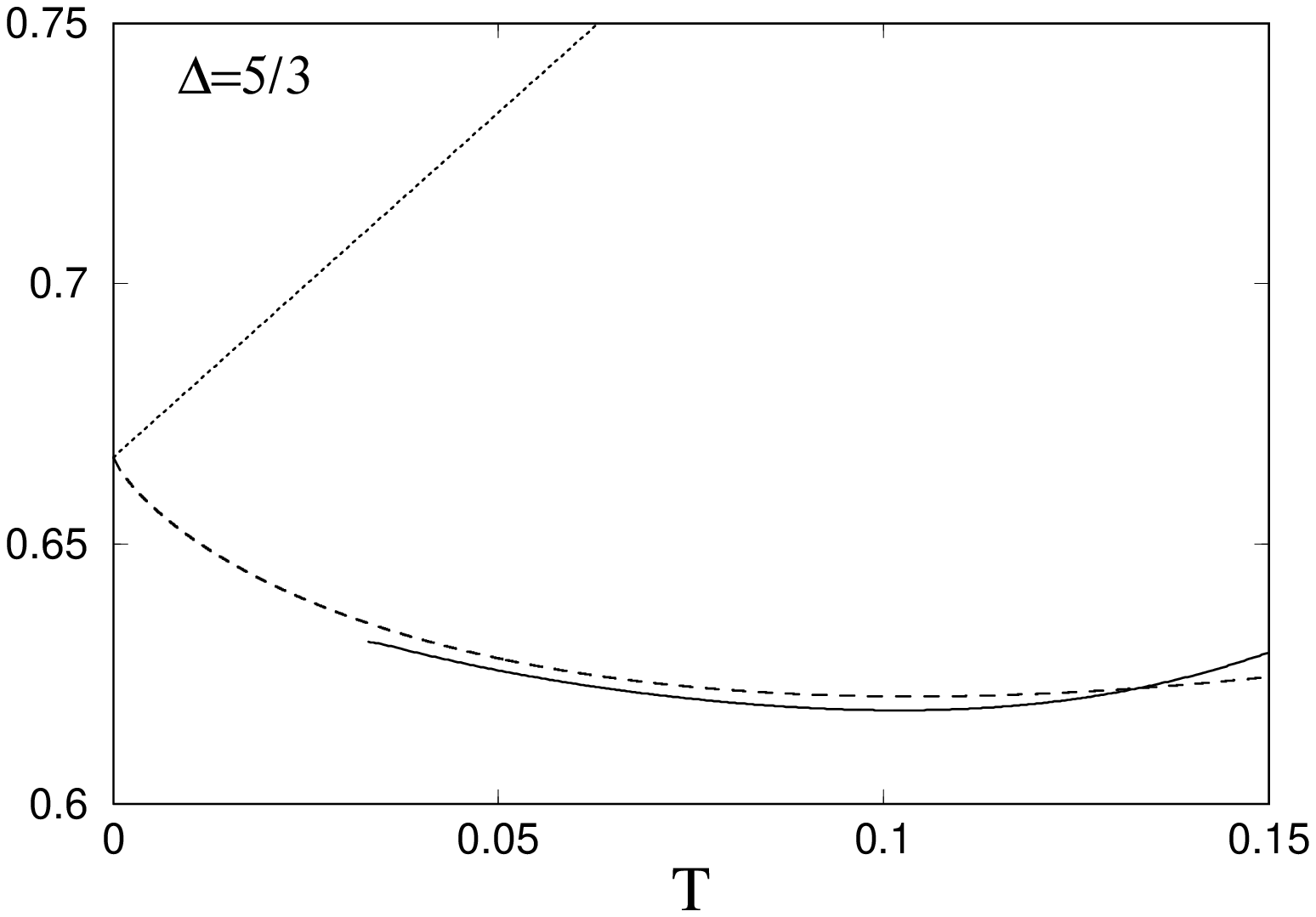,width=5.4cm}}}

\vspace*{0.3cm}

\caption{Comparison of the calculated specific heat
(solid line) at vanishing field $(H=0)$ with the spinwave
approximation (9) (dotted line) and the Johnson-Bonner
prediction (10) (dashed line).}
\end{figure}

\vspace*{-0.1cm}
In order to understand this situation we now invoke an asymptotic
result obtained for the ferromagnetic $XXZ$ chain by Johnson and 
Bonner \cite{7} who predict that the low-temperature specific heat
is more appropriately described by
\vspace*{-0.15cm}
\begin{eqnarray}
\label{eq:10}
&&
C\approx {G^2_-\exp (-G_-/T)\over (2\pi T^3)^{1/2}}+
{G^2_1\exp (-G_1/T)\over T^2},
\nonumber
\\ 
&&
G_1={1\over 2}
\sqrt{\Delta^2-1},
\end{eqnarray}
where a new gap $G_1$ is potentially important. This gap 
originates in bound multimagnon or domain-wall states, including
the notorious factor $1/2$ familiar from earlier discussions of the 
antiferromagnetic Ising and $XXZ$ chains. \cite{8} The two terms in 
Eq. (\ref{eq:10}) may then be referred to as the magnon 
and Ising contributions,respectively. 
The two gaps $G_-$ and $G_1$ become equal at the critical
anisotropy $\Delta =5/3$ and are ordered as $G_-<G_1$ or 
$G_->G_1$ for $\Delta < 5/3$ or $\Delta >5/3$.

Therefore, when $1<\Delta <5/3$, the magnon contribution in Eq. (\ref{eq:10})
dominates for sufficiently low temperature, practically in the region 
$T\ll G_1-G_-\equiv\delta$. For $\Delta =3/2$ one finds that
$\delta =0.06$ and hence the region $T\ll\delta$ is difficult to 
approach by the inherently finite-temperature TMRG algorithm.
This explains the apparent failure of spinwave theory demonstrated 
in the $\Delta =3/2$ entry of Fig. 6. However, when both terms of 
Eq. (\ref{eq:10}) are included, the agreement with our TMRG result is 
obviously very good. The picture becomes more transparent in the 
$\Delta =5/4$ entry of Fig. 6 where the differential gap 
$\delta =0.125$ is greater and thus the region $T\ll\delta$ becomes 
accessible to TMRG, albeit somewhat marginally. Also interesting 
is the result for the critical anisotropy $\Delta =5/3$ shown 
in Fig. 6, where the failure of spinwave theory becomes complete,
whereas our result continues to agree with the Johnson-Bonner
prediction (\ref{eq:10}). Finally we have examined the case of a supercritical 
anisotropy, $\Delta =2$, with a similar conclusion.

At finite (staggered) field our model is not equivalent to the 
ferromagnetic $XXZ$ chain and thus the finite-field results
of Ref. 7 are no longer applicable. We do not know at this point 
how to generalize Eq. (\ref{eq:10}) to account for a staggered field,
especially because complete integrability seems to be lost.
Numerical investigation of this issue suggests that spinwave
anomalies persist in the subcritical region $H<H_b$ while normal 
spinwave behavior is restored for $H>H_b$. In the latter region the 
quantity $-T\ln (T^{3/2}C)$ interpolates linearly to the magnon
gap $G$ shown in Fig. 1.

To summarize, we have presented a reasonably complete theoretical 
description of the thermodynamics of the spin-flop transition for 
Hamiltonian (\ref{eq:1}). Our explicit results would be directly relevant
for the analysis of actual experiments, provided that a 
quasi-one-dimensional magnetic system is found that is described 
by our model Hamiltonian at least approximately. \cite{9} Perhaps
equally important is the overall conclusion that the TMRG method
proves to be reliable even under stringent conditions.

We are grateful to A. Orendacova and M. Orendac for bringing Ref. 7
to our attention and for a related discussion.
This work was completed during a visit of JK at IRRMA.
XW and XZ acknowledge support by the Swiss National Foundation
grant No. 20-49486.96, the Univ. of Fribourg and the Univ. of
Neuch\^atel. We would also like to thank A. Kl\"{u}mper for helpful 
correspondence.

\end{document}